# The Blind Spot of BGP Anomaly Detection: Why LSTM Autoencoders Fail on Real-World Outages

Samuel Oluwafemi Adebayo

Independent Researcher, Network Engineering and Security

June 21, 2025

**Author Note**

Samuel Oluwafemi Adebayo is a certified Network Engineer with professional certifications including HCIE, CCNP, 3x HCIP, and 6x HCIA. His expertise lies in network design, automation, and security, with a focus on large-scale routing protocols and anomaly detection.

There are no known conflicts of interest to disclose.

Correspondence concerning this article should be addressed to Samuel Oluwafemi Adebayo.

**LinkedIn**: https://www.linkedin.com/in/samuel-adebayo1/

**Abstract**

Deep learning has significant potential to make the Internet's Border Gateway Protocol (BGP) secure by detecting anomalous routing activity. However, all but a few of these approaches rely on the implicit assumption that anomalies manifest as noisy, high-complexity outliers from some normal baseline. This work challenges this assumption by investigating if a best-in-class detection model built on this assumption can effectively deal with real-world security events' diverse signatures. We employ an LSTM-based autoencoder, a classical example of a reconstruction-based anomaly detector, as our test vehicle. We then contrast this model with a representative sampling of historical BGP anomalies, including the Slammer worm and the Moscow blackout, and with a simulated 'BGP storm' designed as a positive control. Our experience unveils a blind spot of our model: the model easily identifies the synthetic anomaly of high complexity but invariably fails to identify real-world events that manifest in the form of a "signal loss" (e.g., Slammer, Moscow Blackout) or "low-deviation" (e.g., WannaCry) signature. We demonstrate that the model mistakenly recognizes the abrupt cut-off of BGP updates during catastrophic failures as a signal of extreme stability, leading to reconstruction errors of virtually zero and total failure to detect. We conclude that the characterization of BGP anomalies as high-reconstruction-error events alone is a weak and dangerous oversimplification. Our research provides the data-driven case for why hybrid, multi-modal detection systems capable of identifying both high-complexity and signal-loss signatures are required to enable end-to-end BGP security.

## Table of Contents

Introduction

**The Promise and Premise of AI in BGP Security**

The Border Gateway Protocol (BGP) is the routing protocol that underlies the global Internet, enabling information exchange between tens of thousands of Autonomous Systems (ASes). Despite its central role, BGP was not designed with built-in security attributes to provide an environment in which malicious route advertisements or unintentional misconfigurations can result in large-scale service disruption (Gao & Rexford, 2001). The concept of the Resource Public Key Infrastructure (RPKI) does bring in strong security through Route Origin Validation (ROV), but it is not a solution and offers no security against path-based attacks and relies on worldwide adoption. In an effort to fill this security gap, the research community has turned increasingly to Artificial Intelligence (AI), specifically deep learning, in the hope of building intelligent systems capable of learning the complex patterns of normal BGP traffic and actively alerting on anomalous behavior in real-time (Hammood et al., 2022).

This exciting AI application is, however, largely dependent on an unstated hypothesis: that BGP anomalies manifest as high-entropy events. It is generally assumed that attacks such as network worm instabilities or BGP hijacks will inject a burst of noisy, complex, and random signals in the BGP update stream, producing a significant deviation from a learned, stable baseline.

**The Unspoken Assumption**

The premise that anomalies are "noisy" deviations will guide the methodological choices in unsupervised anomaly detection. Based on the premise that anomalies are challenging and disordered, an ideal detection model is one that excels especially at discovering data points that are challenging to fit a pattern to. This has led to the widespread use of autoencoders, a neural network for this very reason. An autoencoder is trained to transform normal data into low-dimensional latent space and back to its original form. As it is being trained only with the benign data, it gets very good at the reconstruction task. It is assumed that when a given anomalous high-entropy sequence is presented, the model will fail to reconstruct it well, and a high reconstruction error can then be detected as an anomaly. Reconstruction-based is a logical and intuitive consequence of the "anomaly-as-complexity" assumption.

**A Tale of Two Anomaly Types**

This research, however, indicates that such an assumption is a false premise. All anomalies are not created equal. Consider two disparate kinds of network events, as depicted in Figure 1:

- The High-Complexity Anomaly: A bungled router configuration or "path hunting" attack might create a wild, random storm of BGP updates, involving high-frequency and rapidly changing AS-paths and announcement and withdrawal. This conforms to the high-entropy assumption.
- The Signal-Loss Anomaly: A large power failure or a network-jamming worm (e.g., Slammer) that makes routers bog down will bring down BGP sessions in its entirety. To a BGP collector, this is not pandemonium; it is a sudden, stultifying quiet—a signal loss.

This binary raises the basic research question: Can a single model, e.g., an LSTM autoencoder built on the premise of detecting high-complexity events, identify also anomalies characterized by the dramatic reduction in complexity?

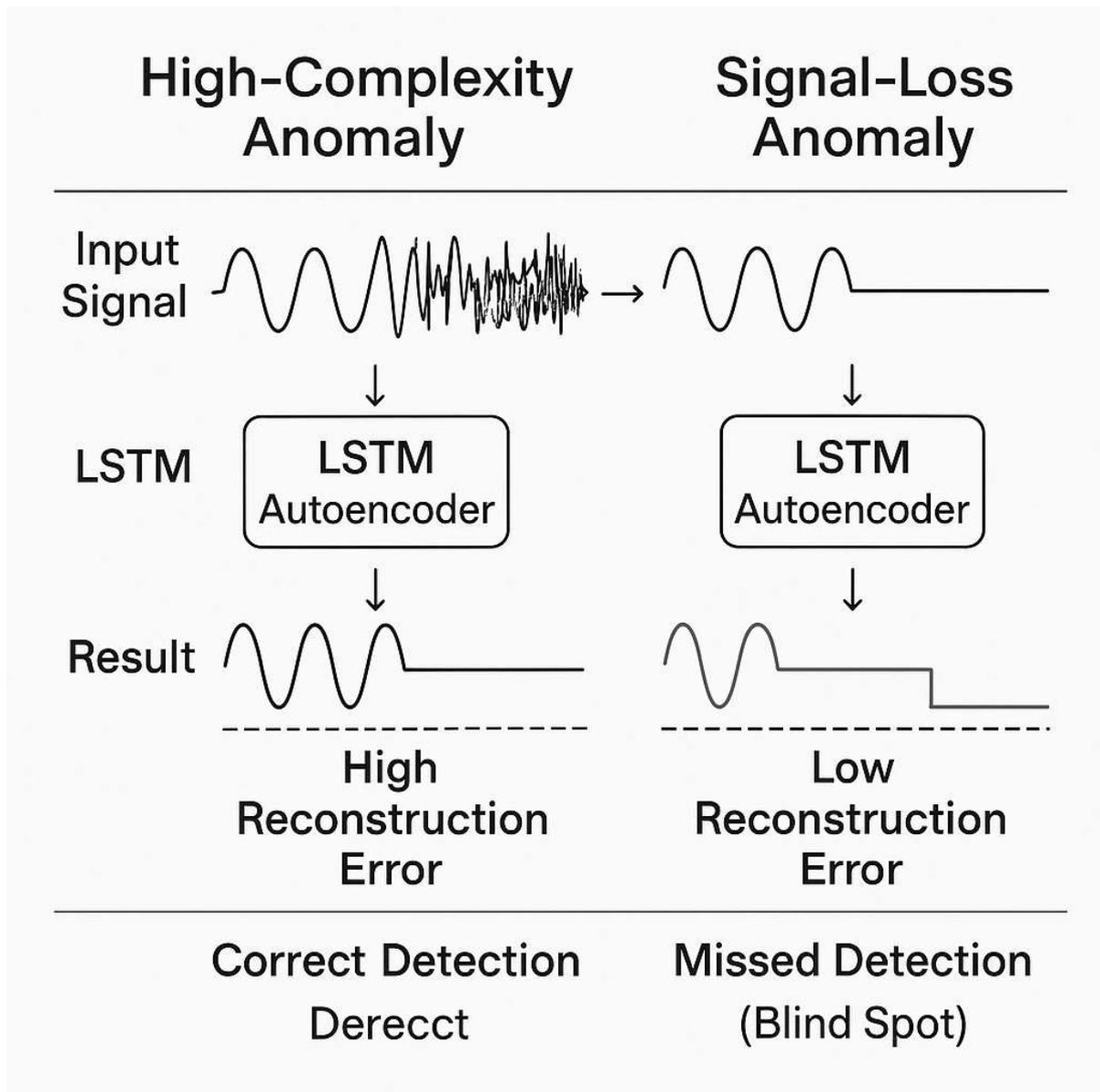

Figure 1 Conceptual Model of High-Complexity vs. Signal-Loss Anomaly Signatures.

**Problem Statement & Contribution**

      The application of deep learning without supervision to BGP security is generally based on the tacit assumption that malicious or abnormal behavior will produce high-entropy, complex signals that are very dissimilar from normal traffic. This research contradicts this assumption by addressing the problem that real-world BGP anomalies come in several different signature forms, which results in catastrophic failure modes in normal anomaly detection models. The main challenge is to demonstrate that a high-performance LSTM

autoencoder, a high reconstruction error-based anomaly detection model, fails consistently when presented with events leading to signal simplification or loss in BGP update streams (e.g., blackouts, session drops) or events with too weak a signature. By validation against examples like the Slammer worm and the Moscow blackout, this paper aims to illustrate that the sole definition of anomalies as high-complexity events is an oversimplification, and a dangerous one at that, and hence provide a data-driven case for hybrid, multi-modal detection systems.

The principal contribution of this work is not to propose yet another detection mechanism, but to provide a critical, empirical analysis that:

- Elaborates on the distinctive BGP signatures of some actual-world anomalies.
- Identifies and establishes a central blind spot in the widespread reconstruction-based anomaly practice.
- Refutes the speculation that all BGP anomalies are "noisy" and shows that some of the most critical outages amount to nothing less than an act of camouflage by simplicity.
- Validates the need for hybrid detection systems that will be able to identify high-complexity as well as low-complexity deviations from the norm.

**Paper Organization**

The remainder of the paper is organized as follows. Section 2 summarizes existing work in machine learning and BGP security. Section 3 describes our system architecture and methodology. Section 4 details the experimental setup, including data and metrics. Section 5 presents our results, including case studies of a number of recent anomalies. Section 6 discusses the broader implications of our results, and Section 7 concludes the paper and outlines directions for future work.

## The Anomaly Detection Paradigm and Its Fault

The machine learning approach to anomaly detection for BGP has advanced significantly, but it has done so based upon a single, overarching paradigm. This section will first explain this paradigm, then speculatively posit a flaw in the underlying assumption of this paradigm, and finally position our work as the first empirical verification of this flaw.

### The High-Reconstruction-Error Paradigm

The current standard for unsupervised anomaly detection of high-fidelity time-series data, like BGP update streams, is reconstruction error. Reconstruction error is based on autoencoders, which are neural networks that include an encoder and a decoder. The encoder projects a high-dimensional input sequence onto a low-dimensional latent representation with its most significant features. The decoder attempts to reconstruct the original sequence from the compressed form.

This is in line methodologically with the theory that anomalies are high-entropy, complex events. The model is only trained on benign data, so it becomes highly specialized in compressing and decompressing the patterns of typical network traffic. The concept is straightforward: when the model sees a sequence following the patterns it has learned, the reconstruction should be correct, and the resulting error should be small. On the other hand, when it is faced with anomalous sequence with chaotic or novel patterns that are new to it, it will find it hard to compress it efficiently, resulting in a low reconstruction and high reconstruction error. This value of error serves as the anomaly score; if it surpasses a statistically derived threshold, an alert is triggered.

A number of papers have been successful in adopting this or analogous principles, using various machine learning algorithms to identify BGP feature set anomalies. Systematic review by Hammood et al. (2020) bears witness to widespread application of these techniques. Studies like that of Li et al. (2020) have proven high classification accuracy feasible on labeled BGP datasets with machine learning models, further solidifying that the detection is a viable option using deviant feature values. The results have cemented the high-reconstruction-error paradigm as a state-of-the-art approach.

**The Theoretical Blind Spot: Anomalies as Signal Loss**

But this model does have one overwhelming theoretical deficiency, which is that it covertly assumes that all catastrophic network failures will come in the guise of increased complexity or noise. We suspect that a major category of catastrophic anomalies does exactly the opposite: it causes blockage of the delivery of BGP updates and, consequently, creates a loss of signal.

Consider the underlying causes of some of the Internet's most disastrous outages:

- Physical Infrastructure Collapse: A severed fiber-optic submarine cable or an extreme power outage (such as that in Moscow in 2005) will physically sever entire data centers or internet exchange facilities.

- Resource Exhaustion Attack: Distributed Denial of Service (DDoS) that overwhelms a network or Slammer-like worm will tax a router's CPU or memory. Under high workload, a router drops fewer critical processes to preserve core services, and preserving BGP sessions active is typically among the first things to be dropped (Labovitz et al., 1998).

In both scenarios, the outcome is identical: BGP peering sessions are terminated. From the perspective of a BGP collector like RIPE's rrc04, traffic from the affected region does not become noisier; it simply disappears. This would be reflected within the feature set as a flow of vectors with predominant entries of zeros or near-zero values for metrics like "number of announcements," "number of withdrawals," and "average edit distance."

We conjecture that for an autoencoder trained on the constant low-level churn of normal Internet traffic, this low-entropy sequence of zeros will be simple to reconstruct. The model will interpret this silence as an extremely simple and repeating pattern, resulting in an extremely small reconstruction error. A catastrophic network crash could therefore be mistakenly identified by the system as a period of excessively good stability. This creates a blind spot of critical magnitude wherein the model is, essentially, incapable of detecting a major class of failures in the real world.

**Placing Our Work**

The above discussion sets up an evident conflict: a commonly used and rational detection paradigm seems to have a theoretical deficiency that would make it blind to particular kinds of extreme anomalies. Though this has been theoretically possible, it has, to our best knowledge, not had a systematic, data-driven falsification against actual historical examples.

The aim of this paper is to be the first to carry out such an empirical falsification. We will counter the high-reconstruction-error paradigm not on an abstract level, but with empirical evidence from the very incidents that demonstrate this "signal loss" effect. By applying a state-of-the-art LSTM autoencoder to historic data from the Moscow blackout and the Slammer worm, we will demonstrate empirically this blind spot. The following sections will describe the experimental setup selected for testing this hypothesis and present the results of its application to a diverse set of historical BGP anomalies.

**Methodology: A Framework for Testing the Blind Spot**

To experimentally test the hypothesized blind spot in the high-reconstruction-error paradigm, we designed a solid experimental framework. In the next few paragraphs, we present the system architecture employed as our test bed, the diverse set of anomalies to be employed in testing, and the rigorous protocol followed to ensure a fair and replicable test.

**System Architecture: The Test Instrument**

To provide a balanced judgment, we first used a state-of-the-art framework that is an archetype of the paradigm we are doubting. The system, as given in Figure 2, is not presented to be a fresh solution but as our primary test device. The LSTM autoencoder framework architecture is designed to precisely perform the reconstruction-based anomaly detection methodology.

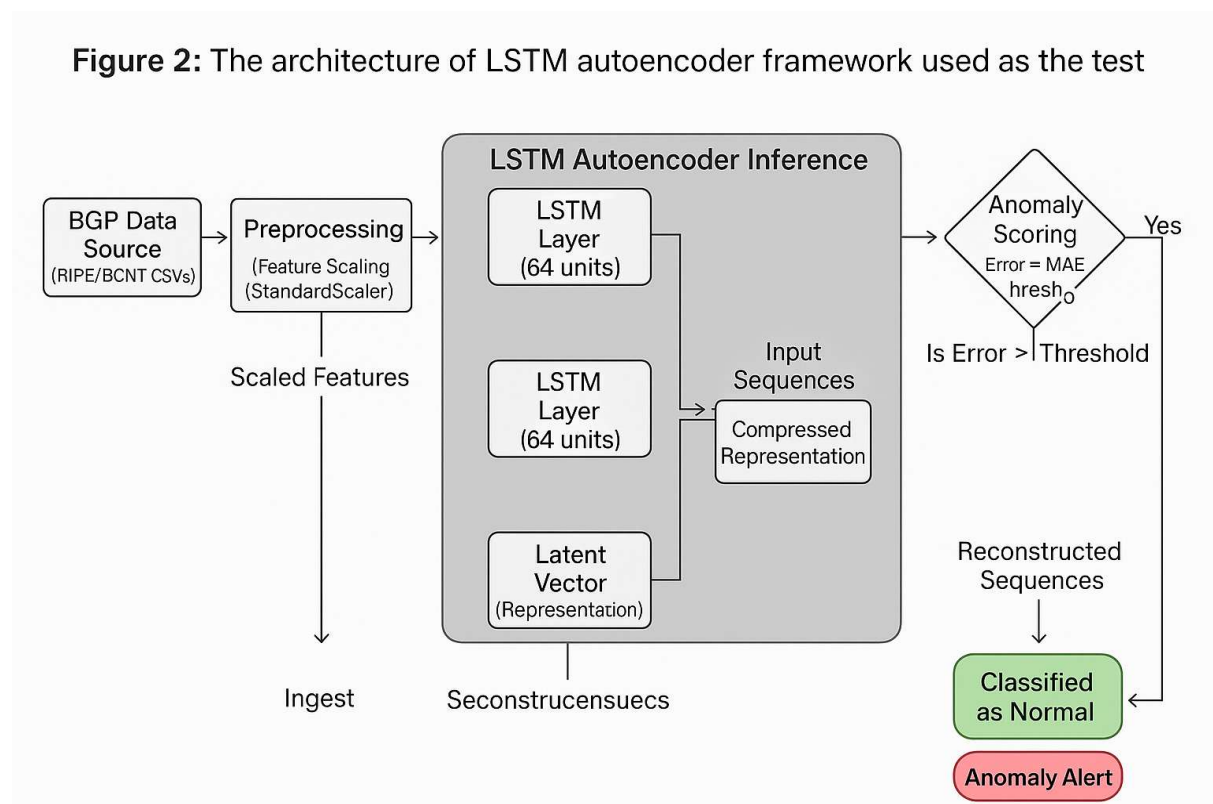

Figure 2 System Architecture of the LSTM Autoencoder Test Framework.

The pipeline consists of five steps:

- Data Ingestion: Loads pre-processed BGP feature data.

- Preprocessing: Rescales the features to unit variance and zero mean to prepare them for use with the neural network.

- Sequence Construction: Rearranges the time-ordered feature vectors into overlapping sequences with a sliding window to build the temporal context for the LSTM.

- LSTM Autoencoder Inference: The core LSTM autoencoder learns from every sequence and attempts to rebuild it from a compressed latent representation.

- Anomaly Scoring & Alerting: Calculates the reconstruction error (Mean Absolute Error) between the input and output sequences. Uses this error as the anomaly score.

By using this standard structure, we are sure that our findings are true for a broad class of such systems and not a product of a flawed or unstandard implementation.

**Datasets: A Diverse Anomaly Portfolio**

The validity of our experiment relies on testing the model against a range of various types of anomalies with highly varying root causes and BGP signatures.

*Real-World Historical Data:* We utilize the public "BGP RIPE Datasets for Anomaly Detection" (Li et al., 2020), containing labeled BGP feature data of five substantial historical events. This dataset presents us with a unique chance to test the model against a collection of real-world events:

- Application-Layer Worm (WannaCry): An anomaly whose impact on BGP may be negligible and difficult to distinguish from normal traffic patterns.

- Infrastructure Failure (Moscow Blackout): Physical power failure predicted to induce a "signal loss" signature as a result of unrestrained BGP session termination.

- Network-Saturating Worms (Slammer, Nimda, Code Red I): Events that caused router CPU saturation, likewise surmised to induce a "signal loss" signature since BGP sessions were dropped under heavy load.

- Benign Traffic: The dataset also contains rich data from RIPE and BCNET that reflects typical operational times, which forms the foundation for training our model.

*Synthetic Positive Control Data:* In order to be certain that our test tool is performing correctly and can recognize anomalies that do reflect the high-complexity hypothesis, we generated a synthetic data set titled "BGP Storm." This data set simulates a wild "path hunting" event, including high-frequency announces and withdraws, extremely long AS-paths, and excessive edit distances. Its purpose is to serve as a positive control—a test case where we should anticipate the high-reconstruction-error method to succeed, thereby confirming our implementation.

**Experimental Protocol**

Our experimental protocol is designed such that it gives a tough and unbiased test of the high-reconstruction-error paradigm.

- Model Training: The LSTM autoencoder is trained only on benign BGP data. This is a strict adherence to unsupervised anomaly detection where the task of the model is to learn an effective representation of "normal" without previously knowing how an anomaly will appear.
- Thresholding Strategy: During training, the model's performance on unseen benign data is evaluated using a dedicated validation set (a chronological split of normal data). The error of reconstruction is calculated for all sequences within this set. Anomaly threshold is fixed at the 99th percentile of this error distribution. This cautious threshold ensures that the system has a very low rate of false positives on regular traffic, a common operational need.

- Evaluation: With the model trained and threshold decided, the various anomaly data sets are executed. For each sequence, a reconstruction error is generated by the model. If it is more than the threshold, the sequence is identified as anomalous.

This protocol ensures an even test by using the paradigm as it must be used: train on representative data and alarm on significant deviations. This configuration being tested against our range of real and simulated anomalies, we can quantify objectively its performance and expose its operating blind spots.

## Experimental Results: Reveal the Blind Spot

This section presents the empirical results of this study. We start by establishing a positive control to verify our test device operates against the right kind of anomaly. We then demonstrate show-case studies with realistic data from real events utilizing historic data, exemplifying the critical blind spots of the high-reconstruction-error detection paradigm when faced with non-high-complexity hypothesis-based anomalies.

**Positive Control: Detection of the Correct High-Complexity Anomaly**

To ensure our LSTM autoencoder and experimental procedure were correctly implemented, we first tested them against our synthetic "BGP Storm" dataset. The dataset was designed specifically to mimic a high-entropy anomaly with chaotic, large-magnitude feature deviations—the exact kind of event the reconstruction-based paradigm is designed to detect.

The results, shown in Figure 3, confirm our hypothesis. The reconstruction error distribution is different for the BGP Storm (red) and is readily distinguishable from and significantly larger than that of typical data (blue). Almost all the anomalous sequences created reconstruction errors much higher than the anomaly threshold. This produced a Detection Recall of 0.9815, since the model successfully detected well over 98% of the anomalous sequences known to be present. This positive control outcome validates that our methodology is functional and valid in the instance where the signature of the anomaly aligns with the "anomaly-as-complexity" assumption.

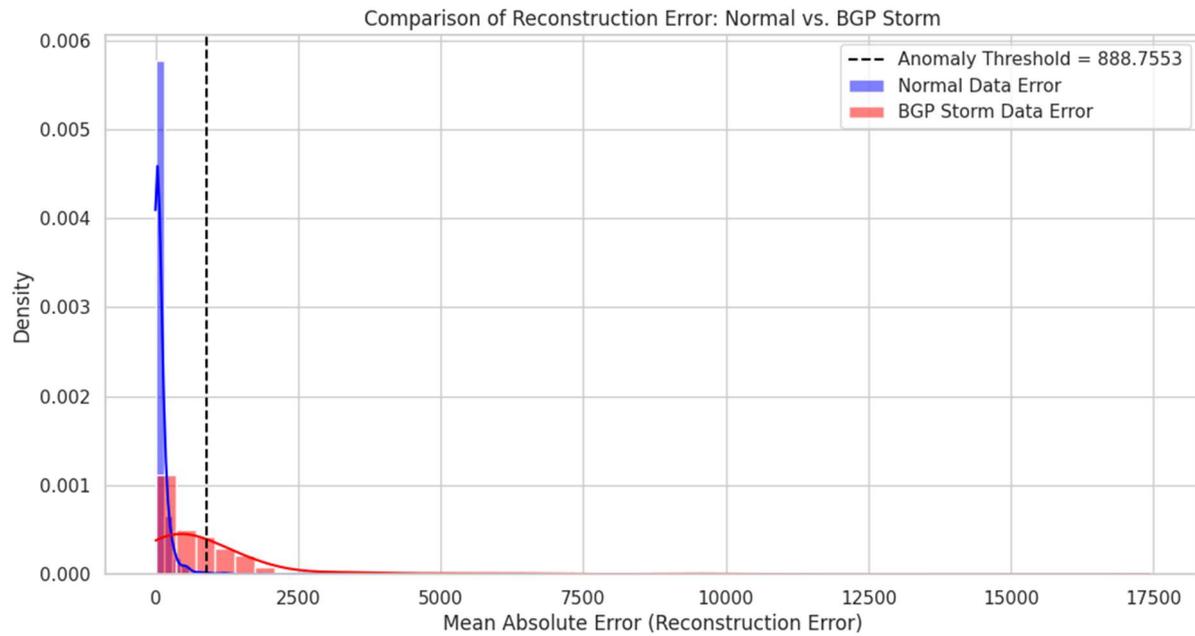

Figure 3 Reconstruction Error Distribution for the Synthetic "BGP Storm" Anomaly.

**Case Study 1: The Signal Loss Blind Spot (Saturation & Infrastructure Failure)**

Having verified our test tool, we next tested it against real-world anomalies suspected to carry a "signal loss" signature. Figure 4 is the result for the 2005 Moscow Blackout (an infrastructure failure) and the 2003 Slammer worm (a network saturation attack).

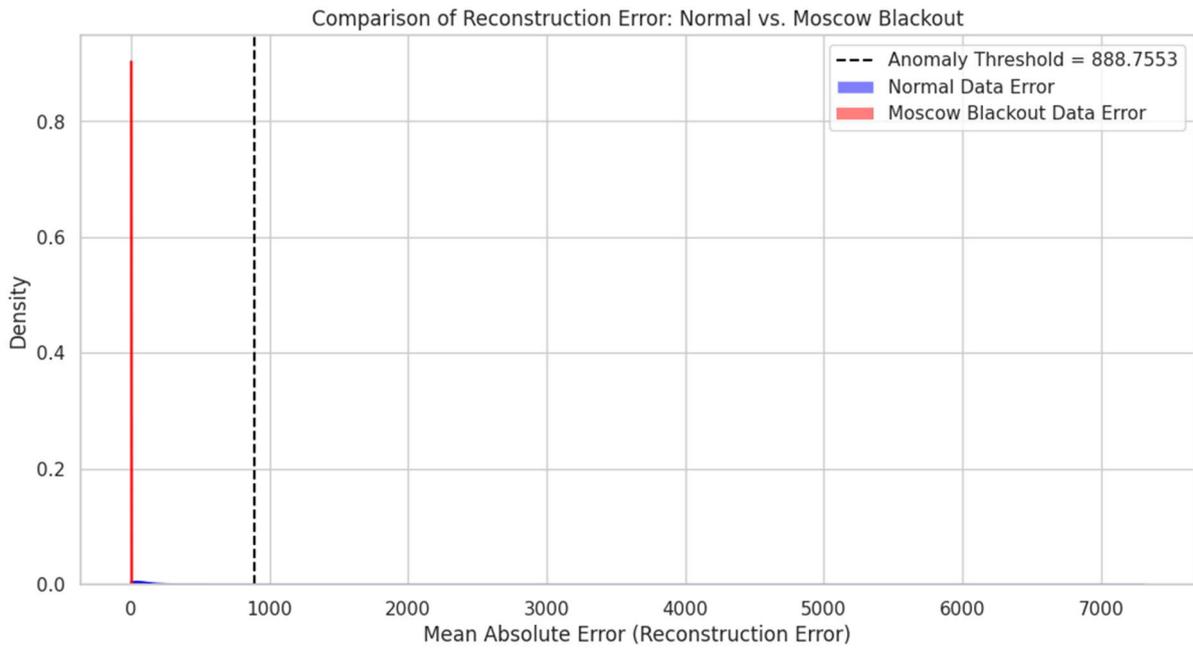

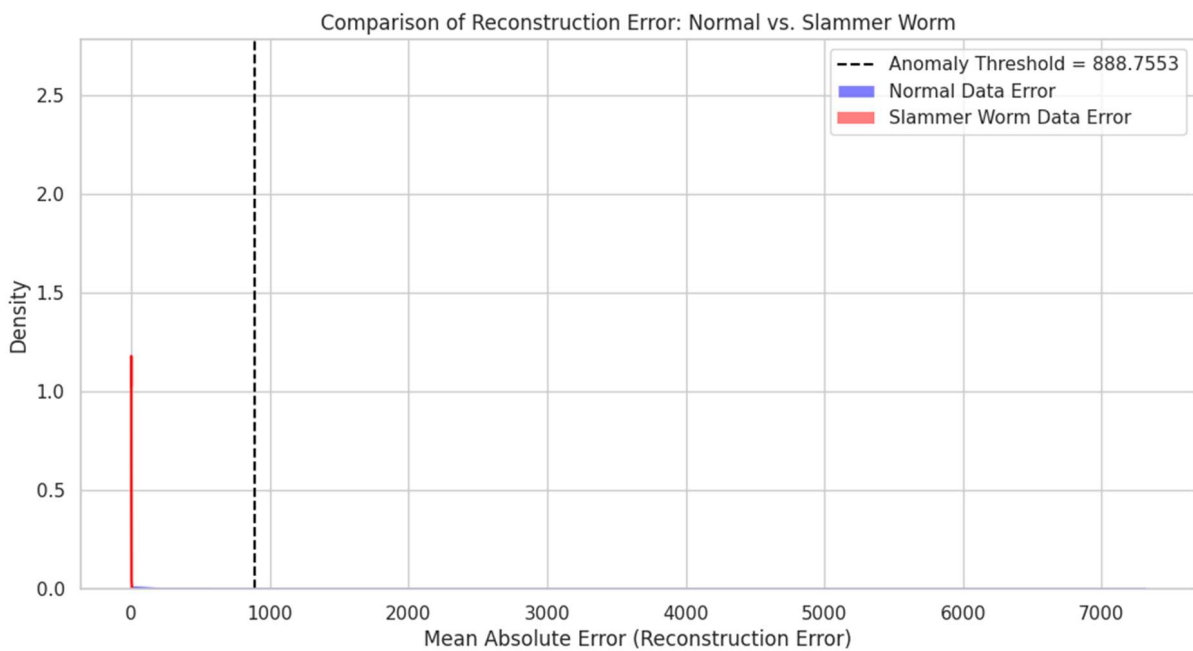

Figure 4 Reconstruction Error Distributions for Signal-Loss Anomalies (Moscow Blackout and Slammer Worm).

The findings are strong and uniform for both events. The reconstruction error distribution of both the Moscow Blackview and Slammer worm are shown as very thin spikes at around zero values—below the median value of standard traffic. Consequently, these events were undetected by the model, and therefore the Detection Recall for both anomalies is 0.0000.

This provides strong empirical evidence for our conjectured blind spot. For the Moscow Blackout, physical loss of power caused stoppage of BGP updates from the affected region. For Slammer worm, network saturation caused routers to drop BGP sessions, which in turn caused loss of BGP signal at the collector. Our LSTM autoencoder, trained to find out complex abnormalities, interpreted the sudden and profound quietude as a moment of unprecedented stability. It was capable of recognizing the resulting sequences of near-zero feature vectors as reconstructable, yielding the near-zero error scores and total detection failure.

**Case Study 2: The Low-Deviation Blind Spot (Subtle Anomalies)**

Lastly, we analyzed a third anomaly type: an application-layer worm whose BGP signature might be too nuanced to cross a conservatively determined threshold. We ran it on data for the 2017 WannaCry ransomware attack. The result is Figure 5.

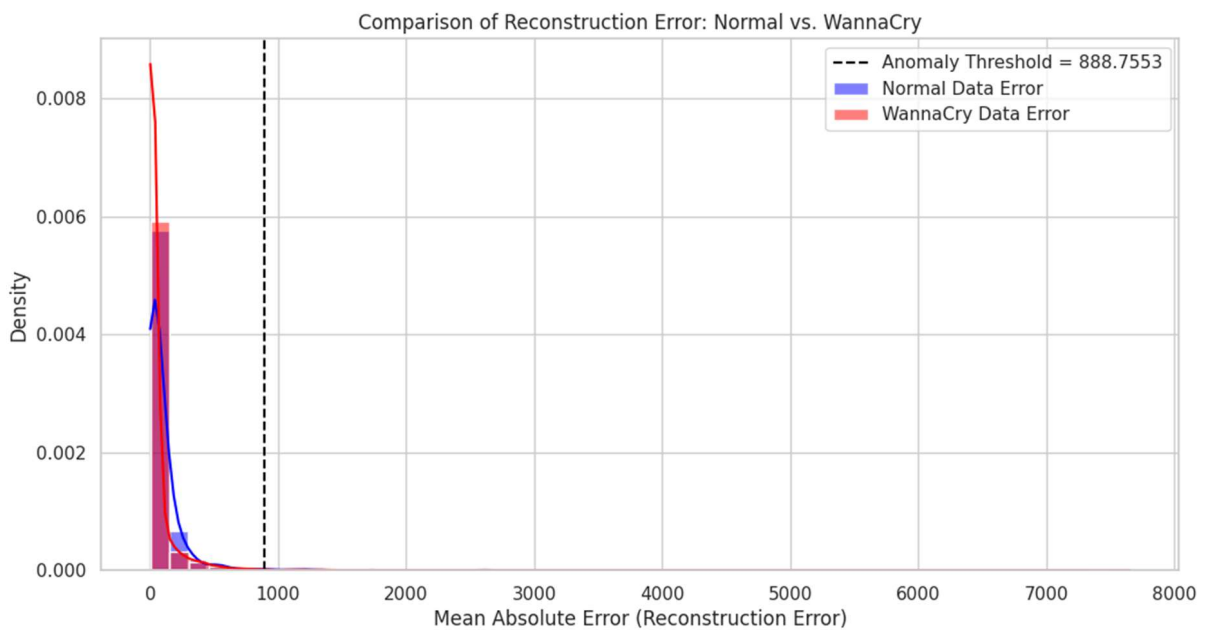

Figure 5 Reconstruction Error Distribution for the Low-Deviation "WannaCry" Anomaly.

WannaCry has a distinct but similar severe failure mode. Reconstruction error distribution of the WannaCry data overlap heavily with the normal data distribution. Though not as direct as the "signal loss" anomalies, its BGP signature was no more dissimilar from the patterns of regular network churn to cause a high reconstruction error. This produced a Detection Recall of just 0.0078, or that the model detected fewer than 1% of the anomalous sequences.

This demonstrates a second blind spot: the low-deviation anomaly. The algorithm is inoperable against attacks whose impact on this specific collection of 37 BGP attributes is too negligible. The conservative threshold, needed to prevent an intolerable volume of false alarms from normal network noise, simultaneously makes the system unresponsive to these weaker deviations.

All three case studies empirically demonstrate that an exclusive reliance on the high-reconstruction-error paradigm for detecting BGP anomalies is insufficient. The system performs excellently in ideal scenarios but is unaware of at least two general types of actual network failures: those that manifest as a loss of signal and those with too faint signal.

## Discussion

The experimental results in the previous section narrate a simple and consistent story: the performance of a reconstruction-based anomaly detection system is solely based on the signature of the anomaly. Our best-available test device behaved as expected on a synthetic high-complexity anomaly but consistently failed on multiple real-world historical events. This section synthesizes these results, proposes a more complete taxonomy of BGP anomalies, and discusses the broader implications for network practitioners as well as security researchers.

**Synthesizing the Failure Modes: The Two Blind Spots**

Our empirical evaluation has uncovered that the widespread "high-reconstruction-error" approach has at least two significant and distinct blind spots if applied to BGP security.

There is initially the Signal Loss Blind Spot. This was vividly demonstrated by the complete inability to sense both the Moscow Blackout and Slammer worm. In both of these events, catastrophic occurrences—one an infrastructure hardware crash, the other a network saturation attack—failed to produce a noisy, complex BGP signal. Instead, they made BGP sessions crash, leading to a sudden cessation of updates. Our LSTM autoencoder, trained on the chronic ebb and flow of a healthy Internet, perceived this "silence" as one of extreme, calculable simplicity. The resulting near-zero reconstruction errors made these outliers wrongly identified as more "normal" than actual normal traffic, revealing a fundamental flaw in assuming all anomalies are complicated.

Second, we identified the Low-Deviation Blind Spot, evident as in the failure to detect the WannaCry ransomware worm. Unlike the signal loss attacks, WannaCry did produce BGP updates, but the overall fingerprint was not unique enough. The profiles were not unique enough from the oscillations in benign traffic to propel the reconstruction error above a conservatively selected threshold. This highlights a classic trade-off in anomaly detection: a high threshold set so as not to be overwhelmed by false positives resulting from common network noise will by necessity be blind to anomalies that cause no radical deviation in the feature space under consideration.

**Redesigning "Anomaly": A Proposed Taxonomy**

Our findings strongly suggest that the catch-all phrase "BGP anomaly" is too broad to be useful in terms of designing useful detection tools. Aggregating an undercover route leak, an unordered path-hunting storm, and a catastrophic blackout in the same catch-all phrase conceals the reality that they require practically different detection methods. In order to bring clarity to this topic, we propose an elementary, functional taxonomy for BGP anomalies by their detection signature:

- Type I: High-Complexity Anomalies. These are events that conform to the classical assumption. They introduce disordered, high-entropy, or new patterns into the BGP update stream. Examples would include our synthetic "BGP Storm," certain types of route flapping, or complex, multi-origin hijacks. These are the ones that high-reconstruction-error models are best at spotting.
- Type II: Signal-Loss Anomalies. They are events representing an unexpected and sustained decrease in the level of BGP update activity and richness. They are caused by events leading to termination of the BGP session, such as physical outages or running out of router resources. As we have shown, these events call for a detection method that can detect an unusual lack of signal, rather than a surplus one.
- Type III: Low-Deviation Anomalies. They are events whose BGP signature is statistically undistinguishable from normal network churn. They may include low-intensity route leaks, application-layer assaults with a very low-key BGP footprint like WannaCry, or early, low-intensity stages of an even broader attack. Arguably, it is the most challenging job to identify these anomalies, perhaps requiring stateful, multi-source data correlation or very advanced supervised models.

Embracing a more precise taxonomy like this is the starting point to creating improved and more robust security solutions.

**Implications for Practitioners and Researchers**

The conclusions from our study have severe implications for the network security community.

For network practitioners and operators, the main takeaway is that there is no "silver bullet" detection tool. Deploying a single, unsupervised reconstruction-error-based anomaly detection system into the field is an illusion of security, since it will be completely blind to some of the most significant types of outages. A defense-in-depth security approach must involve several, complementary systems. The ideal BGP security suite would combine building-block mechanisms like RPKI with observer functionalities that are able to detect different types of anomalies—one for high-complexity events, and another for signal-loss events.

Security researchers are invited by this work to move beyond the monolithic high-reconstruction-error paradigm. Future directions for research should include hybrid detection systems. Such a framework might include our LSTM autoencoder (for Type I anomalies) in conjunction with an ancillary statistical model—very plausibly one that monitors the moving average and standard deviation of BGP update volume to detect the "unusual silence" of Type II anomalies. As recognized by previous reviews, the field is advancing, but our findings highlight a specific, pressing direction for that advance (Hammood et al., 2022). The goal should not be to build a better version of the same tool, but to build a toolkit of tools with different instruments designed for different parts of this many-sided issue.

## Conclusion and Future Work

**Conclusion**

The prospect of using AI to lock down the Internet's routing infrastructure has driven extensive research into unsupervised anomaly detection. This research has been empirically demonstrated to establish that the overall approach of utilizing deep learning models in detecting BGP anomalies by high reconstruction error is flawed in itself when it stands alone. Our tests, using a cutting-edge LSTM autoencoder, validated its capability to detect a synthesized, high-complexity "BGP storm," verifying the reliability of the approach under ideal conditions.

But when tested against a portfolio of history, real-world events, blind spots in the system were exposed. We have shown this paradigm to regularly miss:

- Signal-Loss Anomalies: incidents like the Moscow Blackout and network-saturating Slammer worm that make BGP sessions lost and lead to a deceptive "silence" within the data stream.
- Low-Deviation Anomalies: Subtle threats like the WannaCry worm whose BGP signature is not sufficiently different from normal network churn to cross a conservative detection threshold.

The key finding of this work is that the characterization of BGP anomalies as solely high-entropy events is a dangerous oversimplification. The failure of such widely accepted paradigms to detect some of the most severe types of network outage guarantees that such a system, employed in isolation, will never be capable of providing end-to-end BGP security.

**Future Work: The Road to a Hybrid System**

The clear implication of our findings is that one, monolithic detection system is not enough. The way forward for BGP security research is the design of hybrid, multi-modal detection systems that are specifically designed to handle the many forms of anomaly that we have characterized. We foresee an architecture in the future that brings together multiple specialist modules:

- Module 1: High-Complexity Detector (For Type I Anomalies): Even an LSTM autoencoder, as employed here, remains a valuable tool. Its proven capacity for detecting chaotic, high-entropy events makes it the first choice for the detection of anomalies such as BGP path hunting or high-intensity route flapping.

- Module 2: Signal-Loss Detector (For Type II Anomalies): To fight the most detrimental blind spot, there must be a complementary module to identify an abnormally missing signal. This could be accomplished with statistical process control methods or change-point detection procedures. A second, less complex but effective approach would be a statistical "heartbeat" monitor that tracks the moving average and standard deviation of BGP update volume of important peers. An abrupt and sustained fall below a traditionally normal floor would trigger a Type II alert, in effect turning the system's current vulnerability into a new detection threshold.

- Module 3: Signature-Based Detector (For Type III Anomalies): Low-deviation, subtle anomaly detection remains the most challenging to break through. This will likely involve moving beyond strictly unsupervised methods. A promising direction is to develop a supervised module with signatures for known subtle attacks or specific policy violations. This would more closely resemble a traditional Intrusion Detection System (IDS), generating high-fidelity alarms for a particular set of threats.

With the integration of these specialized modules—a noise-listening module, a silence-listening module, and a known-fingerprint-search module—a highly robust and resilient BGP security system is possible. Further work should be geared toward designing, implementing, and integrating such a hybrid system.